**Evidence-Based Education and Beyond: The Critical Role of Theory in Science Education Research and Practice**

Christoph Kulgemeyer[a]*, Anna Weißbach[b], Kasim Costan[c], David Geelan[d], David Treagust[e]

[a]* Corresponding author: Christoph Kulgemeyer, Institute of Science Education, Physics Education Group, University of Bremen, Otto-Hahn-Allee 1, 28359 Bremen, Germany, E-mail: kulgemeyer@physik.uni-bremen.de, https://orcid.org/0000-0001-6659-8170

[b] Institute of Science Education, Physics Education Group, University of Bremen, Germany, anna.weissbach@uni-bremen.de

[c] Institute of Science Education, Physics Education Group, University of Bremen, Germany, kasim@uni-bremen.de

[d] School of Education, The University of Notre Dame Australia, Sydney NSW 2000, Australia, david.geelan@nd.edu.au

[e] STEM Education Research Group, School of Education, Curtin University, Perth, WA, 6102, Australia, D.Treagust@curtin.edu.au

**Running Title:** Role of Theory in Science Education

*Statements relating to ethics and integrity policies*
- data availability statement: not applicable (theoretical study)
- funding statement: no third-party funding
- conflict of interest disclosure: the authors report no conflict of interest
- ethics approval statement: not applicable (theoretical study)
- patient consent statement: not applicable (theoretical study)
- permission to reproduce material from other sources: not applicable
- clinical trial registration: not applicable (theoretical study)

**Acknowledgements:** The authors would like to thank Professor Horst Schecker (University of Bremen) for his valuable insights, feedback, and contributions during the preparation of this manuscript.

**Evidence-Based Education and Beyond: The Critical Role of Theory in Science Education Research and Practice**

## Abstract

Evidence-based education has become a central concept in science education, with meta-analyses often regarded as the gold standard for informing practice. This emphasis raises critical questions concerning the applicability, generalizability and transferability of research findings into classroom practice. It remains unclear both what kind of evidence education should be based on and whether science education research can provide the type of evidence required to guide decisions at different levels. This paper argues that theories play a crucial role in building bridges between research and practice. Drawing on literature from science education and the philosophy of science, we contrast the explanatory scope of meta-analyses with the predictive and integrative potential of theories, understood in a structuralist sense as systems of models with defined domains of applicability. We propose that science education research requires both fundamental and applied research, each contributing to theory development at different levels, ranging from local and context-specific models to more fundamental theoretical frameworks. Importantly, we argue that theories in science education should not be viewed merely as applications of psychological or pedagogical theories, but as fundamental theories in their own right. We conclude that the future development of science education research may benefit more from the systematic refinement and integration of theories than from the continued accumulation of isolated local findings, and we propose ways to support the development of such theories. A theory-guided understanding of evidence-based education can strengthen the scientific foundations of the field while simultaneously enhancing its practical relevance, thereby helping to narrow the long-standing theory–practice gap.

**Introduction**

Evidence-based education has become increasingly important. Scientific evidence is expected to inform decisions at very different levels, ranging from political decisions that shape educational systems to instructional decisions made by individual (science) teachers. However, it is often unclear what kind of evidence is being referred to in this context—for example, whether it consists of findings from individual empirical studies or from meta-analyses. From a meta-perspective, it is not even certain whether educational research—and science education research in particular—can, in principle, provide the type of evidence required to guide decision-making at all these levels.

We argue that a closer examination of the comparatively recent emphasis on evidence-based education requires revisiting an older, yet crucial, discussion: what "theory" can mean in the context of science education and how it relates to both research and practice. As has been argued previously (Suppes, 1974), theories shift the focus from merely empirically documenting effects to investigating why these effects occur and how they emerge. Theories are important for teachers and researchers because they make educational events and processes understandable (Gage, 2009, p. 39). More fundamentally, reflection on the role of theory lies at the core of any scientific discipline, and science education research, in particular, may benefit from engaging more explicitly in this discussion. A central question in this context is whether science education can develop its own domain-specific fundamental theories or whether it must rely primarily on theories from related disciplines such as psychology, pedagogy, or the natural sciences.

Against this background, the present paper pursues two closely interconnected goals: (1) to stimulate discussion among science education researchers about the role of theory in the field, and (2) to examine what evidence-based education can mean for science education. We propose an understanding of evidence-based education that is explicitly guided by theory. To support this aim, we adopt a structuralist perspective on theory, which has proven fruitful for understanding theories in disciplines such as psychology and may also be beneficial for science education research. The purpose of this paper, however, is not to persuade researchers to adopt our perspective; rather, it is to invite counterarguments and promote discussion.

In one of the few recent reflections on the role of theory in science education research and practice, Svoboda, Sevian, and van Dusen (2025) rightly point out in a paper published in *Science Education* that theories shape "what researchers notice, value, measure, and attempt to explain" (Svoboda, Sevian, & van Dusen, p. 3) and call for more explicit reflection on theory. Indeed, discussion of the role of theory has not been particularly vivid in recent years, and consequently many of the references we draw upon in this paper are relatively old. We argue that revisiting these contributions can foster an important meta-level reflection on the development of science education as a scientific field. In the present paper, we argue that:

1. A more nuanced engagement with theory in science education could substantially enrich current debates on evidence-based education.

2. Theories are crucial for science education practice because, contrary to common assumptions, they are often more applicable and informative for practice than meta-studies.

3. Theories are essential for science education research to grow as a scientific discipline. However, studies in this field currently engage only rarely in the systematic refinement and extension of existing theories.

4. Science education research too often fails to develop coherent theories that organize phenomena through generalizable principles and allow predictions of educational outcomes. Instead, the term theory is frequently used in a superficial sense—as a collection of study results (in the sense of collective pedagogical content knowledge), methodological guidelines, or conceptual frameworks rather than as a structured explanatory system.

    5.  Both fundamental and applied research, using both qualitative and quantitative data, are necessary to develop theories in science education.

 Given the broad scope of science education, our discussion focuses on one of its core objectives: understanding student achievement. We acknowledge that science education also addresses, e.g., normative, sociological, and epistemological questions. Moreover, the research paradigm clearly matters, as it influences whether and how theories are regarded as important (Treagust & Won, 2023). However, this focus on the core objective of the research focused on understanding student achievement is necessary to sharpen the conceptual framework we propose.

**Framing the problem**

 Over the past two decades, evidence-based education has increasingly been promoted as the gold standard in education (e.g., Slavin, 2020; Kiemer & Kollar, 2021, Nelson & Campbell, 2017). The analogy to evidence-based medicine is evident: in medicine, practitioners are expected to act in accordance with empirically established guidelines, and those who deviate from them must provide a strong justification. This parallel raises an obvious question: why should education—and science education in particular—be any different? Both in education and in medicine, practitioners must make decisions based on their reflection on a given situation, with scientific evidence playing an important role in informing these decisions. Of course, the state of research differs. Medicine with its centuries-long tradition has developed a robust body of knowledge that enables evidence-based decisions across many contexts. Science education, by contrast, still requires many more years of research to reach a comparable breadth and depth of knowledge as medicine has today. Some have argued that medicine has undergone a shift from eminence-based to evidence-based decision-making (Isaacs & Fitzgerald, 1999), in the sense that practitioners no longer rely primarily on the authority of senior colleagues—the "eminences" of the field—but instead base their decisions on scientific evidence. Education is also moving in this direction, though in a less linear manner, partly due to different philosophies in relation to the goals of learning—be that inquiry learning or direct instruction, both of which can have their place in the classroom—but also because of the role of politics. One problem in education that seems not to occur in medicine is that politicians, often without any educational qualifications, promote a particular agenda for learning that is often negated in the future after a change of government.

Nevertheless, aiming for medicine as a role model in this regard seems like a worthwhile goal. But what does *evidence-based education* actually mean? Does it imply that empirical studies guide instructional decisions? If so, this raises several challenges: effect sizes are often small, and in many cases one may question whether effect sizes are large enough to be meaningful in a school context. They may be relevant for informing decisions at the level of educational systems, but not necessarily for guiding the decisions of individual teachers. Also, studies sometimes yield contradictory results. The debate around the so-called "replication crisis" has revealed not only a lack of replication studies in general but also that replication attempts frequently fail to confirm original findings (even though over the last few years positive changes have been reported, for example, see e.g. Korbmacher et al., 2023).

Meta-analyses or critical reviews represent one approach to addressing these issues and may therefore constitute a more appropriate source of evidence for informing evidence-based teaching than single empirical studies. Does this mean teaching should be guided by meta-analyses? For example, in the context of the very influential work of Mitchell (2014) or Hattie (2009), this meaning is suggested. That may be preferable—assuming the meta-analyses are conducted rigorously. Yet even meta-analyses typically address only very specific aspects of teaching, for example comparing different approaches to integrating laboratory work in science education (e.g., Hofstein & Lunetta, 2004). What about other aspects? Surely, we should not postpone teaching them until research has advanced far enough for meta-analyses? Furthermore, while meta-analyses sometimes provide clear recommendations under specific conditions, they rarely yield universally applicable conclusions. For example, the finding that direct instruction is highly effective for students with special needs (as suggested by Mitchell, 2014, p. 174) or even that direct instruction is more effective than inquiry-based learning (as suggested by Hattie, 2009, p. 204 and p. 209) may hold true in certain contexts and for specific learning objectives, but not in others. For starters, for acquiring content knowledge, direct instruction might be effective but maybe not for learning inquiry skills. Meta-analyses in general cannot offer guidance without specifying the conditions under which their recommendations apply. Moreover, meta-analyses sometimes synthesize studies conducted under very different conditions—for example, across different educational systems—making it unclear whether and how these contextual differences influence the reported outcomes.

This is where theories have the potential to make a contribution. A theory is, broadly defined, "a verbal construct that organizes phenomena within a well-defined domain in a propositional or conceptual way, describes the characteristics of its objects of interest, enables the derivation of general laws for these objects, and allows predictions of phenomena within its scope" (Mittelstraß, 2024, p. 20; translation by the authors). This meaning of theory comes from the philosophy of science and is prominent in the literature on the nature of science in science education as well (e.g. Edelsztein & Vormick, 2023). We argue that it is valuable to broaden the discipline's understanding of evidence-based education by incorporating a stronger role for theory. A well-developed theory of science instruction could, in principle, predict the most effective ways to teach specific content across a much wider range of circumstances than the narrow scope allowed by meta-studies. Crucially, such a theory could also offer guidance for situations that have not yet been empirically investigated but which still fall within its defined domain of applicability. In physics, for example, there is no need to study every possible moving object individually to assume that Newton's laws apply. Of course, theories in physics differ substantially from those in education (as discussed in the section "What is a concept of 'theory' for science education?" below). Nonetheless, we argue that meta-analyses are inherently too limited to provide broad, generalizable guidance for instructional practice, whereas theories have the potential to do so.

Importantly, even from this perspective, meta-analyses remain vital:not as the primary source of directly applicable knowledge, but as essential tools for identifying research desiderata, testing theories, and-- depending on one's concept of theory (see the section "What is a concept of "theory" for science education?" below) -- finding applications for a theory. In this sense, theories serve as a conceptual bridge between meta-analyses and teaching practice. For research, meta-analyses test theories, but theories guide which meta-analyses are conducted and which empirical studies are included. A theory that seeks to describe how science instruction should be designed may generate specific recommendations for teaching content knowledge. However, if these recommendations fail when examined through meta-analytic evidence, the theory must be revised—or even replaced. This is an essential role of theory (Gage, 2009).

But what kind of theory would this be? Perhaps theories like cognitive load theory (Sweller, 1988), self-determination theory (Deci & Ryan, 2008) or conceptual change theory (Posner et al., 1982) could be candidates—but even though they might be particularly relevant for science education, they are not specific to science education. Suppes (1974) influenced many following researchers (e.g., Scanlon, Alfey & Ovens, 2025) by offering several arguments for the importance of theories in education. He emphasized that theories advance research by shifting the focus from merely recording differences between different instructional approaches based on their effect size to investigating *why* such differences occur and *how* they emerge. Moreover, he argued that the mere accumulation of empirical facts impedes the transfer of findings from one context to another, since it remains unclear what these situations share in common—precisely the kind of connection that theory can provide. If we accept the worth of theories in principle, it becomes evident that science education research needs to provide teachers with such theories.

Another issue with evidence-based education must not be overlooked. Teachers sometimes observe outcomes that contradict what meta-analyses suggest; many tend to trust their own or their colleagues' experiences more than meta-analytic findings (Schmidt et al., 2022). For teachers, such discrepancies might pose a problem. When their experiences conflict with meta-analytic advice, it widens the theory–practice gap and may erode trust in science education research over time (Costan et al., 2026). This disconnect may stem from overgeneralizing meta-analytic findings without considering contextual factors. Here again, theory could provide a better foundation for guiding instructional practice, while experience remains essential for applying theories in specific contexts. Interestingly, evidence-based medicine explicitly acknowledges the role of practitioner experience: clinical decisions are informed by both individual expertise and research evidence—a perspective that shares similarities with approaches like action research in education where practice and research collaborate.

**The theory-practice gap and how science teachers perceive evidence from the educational sciences**

Relatively little is known about how teachers perceive science education research. What is well documented, however, is that the transfer of research findings into classroom practice is often unsuccessful: innovative teaching materials, for instance, rarely become established in everyday instruction, even when they have been empirically validated and proven effective (Breuer, 2021). This persistent "theory–practice gap" has been described frequently (e.g., Nägel et al., 2023; Vanderlinde & van Braak, 2010). A further complication lies in the fact that researchers and practitioners often operate with divergent understandings of what educational research entails (Edwards et al., 2007). Within the research community itself, there is disagreement about the nature, aims, and methodologies of the field (Vanderlinde & van Braak, 2010), whereas teachers frequently expect immediate practical relevance and criticize research for being overly abstract or theoretical (e.g., Schmidt et al., 2022).

Teachers' attitudes toward educational research appear to play a decisive role in this process. Vanderlinde and van Braak (2010) reported that teachers perceive research questions as misaligned with practice and research findings as ambiguous or contradictory. Other studies highlight similar concerns: teachers' skepticism toward research knowledge has been shown to hinder the uptake of academic findings (Schaik et al., 2018), and such attitudes have even been identified as the strongest predictor of research use in practice (Lysenko et al., 2014). Some educators express doubt about the value of educational research altogether (e.g., Gore & Gitlin, 2004; Nicholson-Goodman & Garman, 2007; Nägel et al., 2023). This skepticism has also been studied specifically in science education. Costan, Costan, Weißbach, and Kulgemeyer (2026), for example, used latent profile analysis to show that physics teachers can be grouped according to their views of science education research. Fourteen percent of the teachers were highly skeptical about the value of research findings in science education, and in this study, most teachers across all profiles perceived the collaboration between research and practice as largely unsuccessful, in the sense that they felt science education research too often operates with insufficient cooperation with teachers (Costan, Costan, Weißbach, and Kulgemeyer, 2026).

Similar findings exist in the broader field of educational research. Teachers often perceive the relevance of current research results as limited, tend to place greater trust in the experiential knowledge of colleagues than in scientific studies, and sometimes doubt whether researchers are capable of addressing the complex realities of classroom life (Nägel et al., 2023; Lysenko et al., 2014). Even when teachers do not explicitly reject scientific evidence, they may view it as overly abstract and therefore dismiss it as a basis for guiding their own practice. Moreover, science teachers have occasionally been critical of teaching materials that were developed based on research evidence (Breuer, 2021). Taken together, these findings suggest that skepticism toward research—ranging from practical concerns about applicability to more fundamental doubts about its value—represents a key barrier to bridging the gap between research and practice. In the following, we will argue that meta-studies alone cannot bridge this gap – and that theories play a crucial role in supporting evidence-based education.

**Evidence-Based Education Following Meta-Studies: The Gold Standard of Teaching?**

The limitations of meta-analyses have been discussed in greater detail by other scholars (e.g. Renkl, 2022). This discussion is valuable for gaining a deeper understanding of the role of theory in both research and practice. A good example of these limitations of meta-studies is arguably the most well-known meta-analysis in education: John Hattie's "Visible Learning" (2009), a synthesis of multiple meta-analyses that has been widely cited as a catalyst for educational innovation worldwide. Yet even this comprehensive work cannot offer specific guidance for every instructional context. For instance, policymakers frequently cite Hattie (2009) to argue that class size has only a minimal effect on learning outcomes. But the critical question is: under what conditions does this apply? Hattie discusses this issue in considerable detail and concludes: "The reader is reminded that meta-analysis is a method of literature review—the lack of effects from lowering class size and these experiences indicate that reducing class size has not been a powerful moderator on outcomes" (Hattie, 2009, p. 88). He further suggests that one possible explanation is that teachers accustomed to large class sizes may fail to adjust their instructional practices when class sizes are reduced, thereby limiting potential benefits. Indeed, class size matters differently across instructional formats: for lectures, the number of students is far less critical than for student-centered laboratory work. In inquiry-based science learning, where students conduct their own experiments, overcrowded classrooms can make effective instruction nearly impossible for both safety and organizational reasons. Conversely, for teaching conceptual knowledge through structured direct instruction, larger class sizes may pose fewer challenges. Moreover, Hattie's synthesis primarily focuses on *achievement* as an outcome variable, while science education aims at a much broader range of goals—including understanding the nature of science, developing experimental competence, and fostering scientific communication skills. Interestingly, an earlier meta-analysis by Glass, McGaw, and Smith (1981) reached an entirely different conclusion: "There is little doubt that, other things equal, more is learned in smaller classes" (p. 23). In short, one might ask: does such evidence from meta-studies truly provide meaningful guidance for practitioners?

Many scholars regard randomized experimental studies and meta-studies that are based on them as the gold standard of evidence that should impact instructional decisions (Davies, 1999; Slavin, 2002; Robinson et al., 2013). Scientific evidence as a foundation for teachers' decisions is widely regarded as important—not only in science education but also in the broader educational research community (e.g., Slavin, 2020). However, scientific evidence comes in many shapes and forms, and there is rarely a discussion about which type of evidence teachers should rely on. A notable exception is provided by Renkl (2022), who primarily discusses the usefulness of meta-analyses for teaching practice. Renkl (2022, p. 218) reminds us that teachers usually do not base their instruction directly on scientific evidence but rather rely on their own experiences as learners (the infamous dictum "you teach as you have been taught") and as teachers. In addition, they are influenced by various external factors, including social expectations from colleagues, students, and parents, as well as available teaching materials. Renkl (2022) argues against viewing scientific evidence as the sole or superior form of knowledge: "Science does not provide all the detailed and empirically backed-up answers to the myriad of questions that teachers have to answer […] it cannot answer normative questions" (Renkl, 2022, p. 218). First of all, there is a greater emphasis in quantitative studies on what can be easily measurable, with the consequence being that there is less evidence on crucial but less tangible things like critical thinking skills and the skills needed to conduct experiments. Meta-analyses typically summarize studies with high internal validity—often conducted in controlled laboratory settings—but with limited external validity, as they exclude many real-world variables to maintain methodological rigor.

The main issue in relying on meta-studies for instructional guidance, however, lies not primarily in the limitations of science education as a field but in the nature of meta-studies themselves. Renkl (2022) outlines three main reasons why meta-studies should not be the primary source of information for teachers: meta-analytic information is "unlikely to change behavior or beliefs" (p. 220), is "not very clear about what's effective" (p. 220); and offers "just one perspective" (p. 221).

Renkl's first reason draws on well-established research (e.g., Pajares, 1992), showing that abstract information rarely leads to changes in beliefs or behavior. Science education, as discussed above, pursues diverse goals, such as fostering conceptual knowledge, understanding the nature of science, or developing experimental skills. To act on these ideas, teachers would thus need to integrate findings from many different meta-studies such as those in well-established fields of research (e.g., laboratory work), where multiple meta-analyses exist (e.g., Hofstein & Lunetta, 2004; Abrahams & Millar, 2008). Renkl refers to Britt et al. (2018) to argue that synthesizing multiple information sources is highly demanding and potentially overwhelming for practitioners. The example we discussed above on classroom size illustrates this. Moreover, meta-analyses rarely provide concrete examples, offering abstract summaries instead—information that is unlikely to support conceptual change in teachers.

Renkl's second reason concerns the frequent heterogeneity of meta-analytic findings (e.g., Pant, 2014). This argument was already addressed above, but Renkl adds a crucial remark: findings often depend on moderators such as prior knowledge or contextual factors (e.g., school environment, metacognitive skills). As a result, teachers may begin to see research evidence as unreliable, and overreliance on meta-analyses may inadvertently fuel skepticism toward educational research.

Renkl's third reason becomes particularly clear in comparison with evidence-based medicine, where practitioners' experience is recognized as a vital perspective to be integrated with empirical findings (Sackett et al., 1996). The same should apply to teaching: informed decisions require both experience and scientific evidence. Biesta (2007) views scientific evidence not as a directive but as a resource for formulating hypotheses that guide intelligent problem solving in practice. Renkl (2022) points out that meta-analyses tend to focus narrowly on a single instructional goal, while teachers often pursue multiple goals simultaneously. Consider experiments in physics education: if the goal is conceptual understanding, it would be plausible that experiments should produce clear, reproducible results in order to demonstrate a concept convincingly. If the goal is to promote understanding of the nature of science, experiments that highlight uncertainty may be more appropriate (Höttecke & Rieß, 2015). If the goal is to foster interest, student-led experimentation may be better; for content acquisition, a teacher-led demonstration might be more effective.

Another important limitation is that meta-studies usually do not explain *why* a certain method is effective, nor whether an alternative might be more effective. As such, they offer local rather than general recommendations: "[...] meta-analyses usually provide rather narrow information. They leave teachers alone with the task to integrate and reconcile their 'local' recommendations with experiential knowledge and other professional (scientific) knowledge" (Renkl, 2022, p. 222). Renkl (2022) describes theories as *primus inter pares* (first among equals) compared to meta-studies—a conclusion drawn from the inherent limitations of meta-studies. We would argue that theories are not *primus inter pares*, but something fundamentally different: a necessary link between meta-studies and teaching practice. To better understand this claim, we will next explore what constitutes a theory in the philosophy of science and in science education.

**What is a concept of "theory" for science education?**

*Different kinds of theories*

The term "theory" can be understood very differently from everyday meanings to very differentiated notions in science philosophy. In everyday language, the term "theory" is often associated with being detached from practice. Thus, in teaching practice, theory is sometimes just framed as the counterpart of practice in general (see above: the so-called "theory–practice gap,"). In a scientific context, it may be understood as "everything we know in science education" and comprises, in that sense, the "collective pedagogical content knowledge" (cf. Hume, Cooper, & Borowski, 2019) in the field. The "theoretical background" as reported in science education research papers usually has an implicit understanding of theory in a similar way: the relevant studies dealing with the topic of the manuscript are reported in order to develop research questions and sometimes hypotheses. Chambers (1992) identified fifteen usages of the term "theory" in educational research. The meaning ranges from theory in the sense of collective PCK, to models or guidelines for a research process, to a very nuanced meaning following the tradition of philosophy of science in research on the nature of science (Schecker et al., 2018, p. 4).

Indeed, the question of what counts as a theory is not easy to answer. Certainly, there are differences between the natural sciences and the social sciences. The philosophy of science has discussed the nature of scientific theory extensively. However, even within this domain, there are different positions: discussing all of them would require a philosophy of science textbook (and, in fact, such textbooks already exist, e.g., Kornmesser & Büttemeyer, 2020). For the purpose of the present paper, however, we need to define what we understand by the term 'theory' in science education and justify this view based on the philosophy of science. As mentioned above, a theory, in its broadest sense, can be understood as "a verbal construct that organizes phenomena within a well-defined domain in a propositional or conceptual way, describes the characteristics of its objects of interest, enables the derivation of general laws for these objects, and allows predictions of phenomena within its scope" (Mittelstraß, 2024, p. 20; translation by the authors). Mittelstraß (2024, p. 24) makes the important claim that philosophy of science does not distinguish consistently between hypotheses, descriptive laws, explaining laws, laws, and theories. This, indeed, differs from the way theory and law are often understood in science education with a focus on the nature of science (Edelsztein and Cormick, 2023).

Kornmesser and Büntemeyer (2024) describe two different views of theory in philosophy of science: the "statement view" and the "non-statement view" or "structuralist theory concept". (see also Mittelstraß, 2018) The former understands theories as a set of statements expressed in a theoretical language. The language of science, in this sense, consists of two levels: the theoretical language and the empirical language. Theoretical language contains only theoretical terms, while empirical language consists only of empirical terms. Newton's laws qualify as a theory because they use theoretical terms such as *force* or *mass*, which cannot be observed directly (Kornmesser & Büntemeyer, 2024, p. 144). However, empirical laws can be deduced from theoretical laws and used to predict empirical phenomena.

***The structuralist theory concept***

The structuralist theory concept uses "models" to describe theories (it is sometimes also called "model-theoretic")—but not models in the sense of models in the natural sciences. Here, models are rather "examples serving as role models" for a theory, that is, to simplify, the parts of reality that the theory should cover. For example, the Earth–Moon system is a model for Newton's laws because this theory can be applied to predict the movement of these two bodies (Kornmesser & Büntemeyer, 2024, p. 145). That the structuralist theory concept understands theories as a set of models means that theories contain multiple such models. The domain in which a theory is applicable (or rather: useful) is the set of these models. Newton's laws, for example, are useful to describe the movement not only of Earth and Moon but of all other observable bodies (provided they are not too fast or too small, which marks the boundary of its applicability).

This structuralist notion of the nature of theory is potentially very helpful for understanding what theories could provide for science education. It has already been applied to describe theories in the social sciences (e.g., Manhart, 1994) and psychology (e.g.,

Westermann, 1987), and we see no reason why it could not be applied to science education research as well. To understand it more fully, it is useful to contrast this structuralist concept of theory with the notion of theory in critical rationalism (which follows the 'statement' view mentioned above). Critical rationalism is arguably the most influential view of science in the 20th century. It follows Popper's ideas that theories are not inductively built from data (as logical empiricism would claim) but rather take the form of hypotheses, which should be empirically tested and subjected to continuous attempts at falsification. In this view, a theory can never be proven absolutely true; it only gains credibility by repeatedly resisting falsification. Scientific progress thus consists of developing theories that explain as many phenomena as possible and attempting to falsify them as rigorously as possible. Critical rationalism understands scientific theories as sets of statements; theories, therefore, have a linguistic core (Manhart, 1994).

Kuhn's (1962) famous critique of critical rationalism also influenced understandings of the nature of scientific theory. Kuhn observed that no scientific progress in history had ever strictly followed Popper's ideas; rather, new theories were more often established through persuasion by influential researchers or, simply, because a new generation of researchers took over. Manhart (1994) sees this as the starting point for a new conception of scientific theory, because not only logic but also social processes turned out to be important in determining what is accepted as a theory. The structuralist theory concept was developed following Kuhn's ideas, most influentially by Sneed (1971) with a focus on theoretical physics. Theories are not regarded as sets of statements but as mathematical structures (hence the term *structuralist* theory concept). This approach is also not normative in deciding what counts as a scientific theory; rather, it describes theories and, by doing so, enables comparison. The main principles for comparing theories have both structural and empirical components. First of all, theories are based on consistent axioms that define the domain for which a theory aims to provide explanations. A phenomenon to which the axioms can be applied is called a "model."

At its core, a theory does not necessarily have any implications for empirical consequences. However, some theories certainly aim to predict observable phenomena—and this is also true for theories in science education research. A theory needs to be well defined in terms of its scope of application. On a structuralist view, what is tested empirically are the intended applications (models) rather than the theory's core structure. As a result, the domain of application can be revised—expanded or narrowed—when models succeed or fail, while the theory itself need not be abandoned outright. Whenever a new theory is introduced, some examples are needed in which the theory has been applied successfully. These are the first models that define the domain of application of the theory.

Subsequently, new models can be identified based on their "family resemblance" to the initial (role-)models; in the example above, any two-body system can be considered to bear a family resemblance to the Earth–Moon system. It must then be tested whether the theory applies to them. If this can be shown successfully, they become models for the theory, even though that status is not permanent. Manhart (1994) claims that this idea is much closer to what really happens in research: rarely has a scientist deliberately attempted to falsify a theory; it is much more realistic that scientists seek applications for a theory. This is important to emphasize: in the sense of the structuralist theory concept, theories are not falsified, but the set of models for a theory is constantly tested. New models are added, expanding the set. Some models may later turn out not to belong to the set based on new empirical results. That, however, does not mean that the theory itself is falsified, but rather that the set of models decreases—in extreme cases, the set of models may eventually be empty. In that case, the theory is no longer an empirical theory, although its mathematical structure remains (Westermann, 1987).

### *Local theories, emergent theories, and fundamental theories in science education*

Theories in science education are certainly empirical, and if the set of models is empty, the theories are no longer useful. Theories in science education can be compared based on the size of their set of models. For the present paper, we call theories having a set of models that includes just one model "local theories," and theories with a potentially infinite set of models "fundamental theories". There are also theories with a finite set of models that can be described as subsets of another theory. We call them "emergent theories."

Local theories can be a specific kind of emergent theory when their models are a subset of those of a fundamental theory. Emergent theories and fundamental theories vary in the breadth of their application, but all three kinds of theories are needed. One example of a fundamental theory might be "conceptual change theory," an emergent theory within this fundamental theory might explain how conceptual change occurs while learning quantum mechanics, and a local theory might be the emergent theory that describes the cognitive changes in the learning pathway of one particular student (e.g., Petri & Niedderer, 1998). Originally, local theories have sometimes been seen as opposed to fundamental theories, with the former representing a qualitative approach to research and the latter representing a quantitative approach (Schibeci & Grundy, 1987). Schibeci and Grundy (1987) describe local theories as "theories that apply within restrictions such as a school or a school system, but not necessarily to other schools or school systems" (Schibeci & Grundy, 1987, p. 92).

In the sense of the structuralist theory concept, however, there is no opposition between local theories and fundamental theories—especially since the structuralist theory concept merely describes theories and does not rate them. Both are scientific theories that differ only in the size of their set of models. Moreover, local theories can be the starting point for a fundamental theory: the first model explained by the theory might just be the starting point, and other models may later be added based on family resemblance and empirical research testing whether they fit into the set of models of the theory. Thus, theories can become more fundamental when they collect new models. This shows that both qualitative research (aiming at local theories) and quantitative research (aiming to test whether a given model falls into the set of models of a theory) are crucial for science education research. While there are notable theory-driven programs in science education, from our experience, systematic efforts to test and refine overarching theories remain comparatively limited, with many studies proposing their own, newly developed frameworks that are rarely integrated into broader theory development. We would also like to point out that action research has a role in this context. A mutual, evidence-guided development of teaching materials by practitioners and researchers may be a powerful tool for bringing innovations into practice (Laudonia, Mamlok-Naaman, Abels, & Eilks, 2018). At the same time, action research might lead to the development of local theories explaining why these innovations work. Activities seeking to combine local theories of this kind to form more emergent and even fundamental theories may not be within the scope and interests of the original action research team, but have the potential to enhance science education more broadly.

**Consequences for an evidence-based science education practice**

*From meta-studies to practice in science education? A naïve heuristic*

"For knowledge users, theory has the inestimable value of making events or processes understandable" (Gage, 2009, 39). Indeed, that is of great value for science education practice. We have argued above that meta-studies do not represent a superior source of evidence for teachers seeking to inform evidence-based practice. Teachers themselves are often critical of the results of science education research and, in some cases, even exhibit a kind of skepticism toward its practical value (Nägel et al., 2023; Costan et al., 2026). We suggest that theories, as outlined, have the potential to bridge the gap between research and practice. Another point worth considering is that if science education does not develop theories that provide generalized explanations of phenomena, teachers tend to construct their own personal theories about what works and why — which can lead to problematic consequences. This has been described in research on "educational myths" (e.g., Vester's learning types), which seem to be prevalent among teachers worldwide (De Bruyckere et al., 2020).

As a heuristic, Figure 1 illustrates the common mechanism that is from our experience often implicitly assumed in discussions of evidence-based education: meta-studies directly informing teachers' instructional decisions. Hattie's famous review has often been discussed like that. We contend that the limitations inherent in meta-studies render this mechanism inadequate and, in fact, potentially counterproductive, as it may encourage the very skepticism toward research that teachers sometimes display. As mentioned above, at their best, meta-studies describe what works under specific conditions. As Renkl (2022) has argued, synthesizing multiple sources of evidence is cognitively demanding and potentially overwhelming for practitioners. We would extend this argument by suggesting that such complexity may lead to overgeneralizations, such as the assumption that "method A is superior to method B under all circumstances." This may explain why teachers sometimes perceive their own experiences as contradicting research findings, thereby reinforcing skeptical attitudes toward science education research.

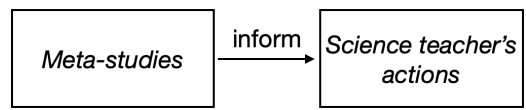

*Figure 1: Evidence-based education based on meta-studies? A naïve heuristic*

### *Refining the heuristic by integrating theories*

Theories of science learning, by contrast, could help to bridge the theory–practice gap because they may predict learning outcomes across a much broader range of contexts than meta-studies and also for contexts that have not been researched yet based on their family resemblance with models for the theory. Figure 2 presents an alternative heuristic, which admittedly still simplifies matters but illustrates the basic relationships and "main effects" among the relevant elements. In this model, theories guide teachers' actions, while meta-studies and quantitative studies test theories—or more precisely, accumulate applications that can serve as models for a theory. Theories also generate research questions and thereby motivate both qualitative and quantitative studies, including meta-studies, by highlighting where further empirical work is needed to strengthen or challenge them. Following the model of evidence-based medicine, teachers' professional experiences remain crucial for informing instructional decisions.

Experience can help determine which theory is most applicable to a given instructional problem. Most interestingly, theory plays a crucial role here as well: The mechanism to turn experience into knowledge is reflection (e.g., McAlpine et al., 1999). It has been argued that reflection requires a theory-based analysis: theory provides the criteria that drive the reflection about experiences. Only if theoretically sound criteria for reflection exist is it likely to be a fruitful process (Kulgemeyer et al., 2022): „Teachers can reflect independently on their own by using theory as an analytical tool in this activity. In this way teachers can use theory both to analyse and understand their practice, to see possibilities in the practice, and furthermore, to improve their practice". (Postholm, 2008, p. 1721).

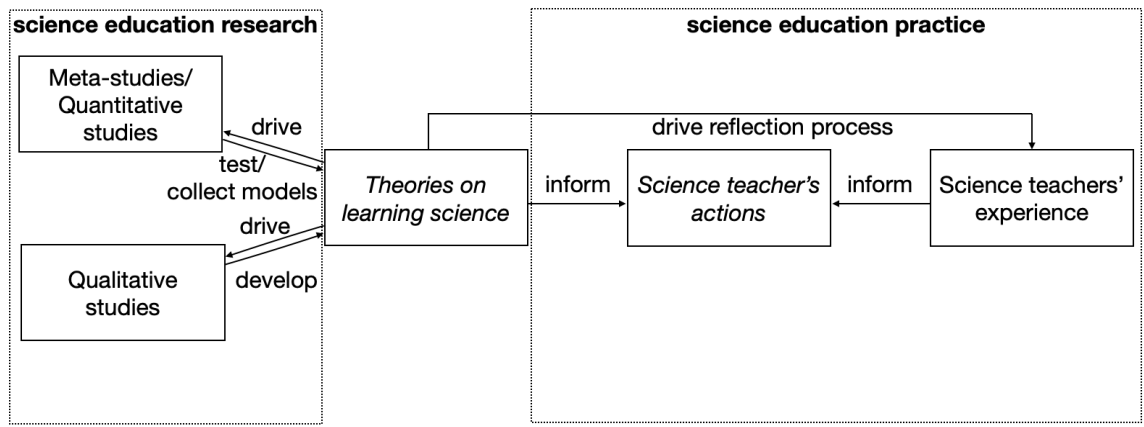

*Figure 2: Evidence-based education based on theories*

On this view, the role of research in science education shifts from providing direct prescriptions for practice to developing and testing theories. Qualitative research is particularly well suited to theory development (as it provides local theories), while quantitative research is valuable for systematically testing theories. Of course, qualitative research can also test theories, and quantitative studies can contribute to theory generation, but these are not their respective primary strengths. Quantitative research offers the potential for generalization (which means testing factors determining family resemblances of models), while qualitative research is indispensable for exploring new or under-researched terrain. Moreover, because quantitative studies often employ standardized conditions (e.g., focusing on specific age groups or school types), they are especially useful for collecting models that can be incorporated into broader theories. In this framework, theories are central to both research and practice, driving each of these processes.

### *What kind of theory does science education practice need?*

This raises the critical question of what kinds of theories are needed—and whether science education research is currently equipped to provide them. At present, theories so fundamental that they could predict learning outcomes across all forms of science teaching are not available. One might even argue that science education has thus far failed to develop such overarching, domain-specific theories. It seems unlikely that a single, all-encompassing theory could capture the diversity of topics addressed in science education, even if restricted to formal schooling. The more pressing question, therefore, is how generalizable theories in science education can and should be.

As a starting point, theories addressing recurring standard situations in science teaching may prove useful. For example, the current state of research might justify the development of theories concerning how to use experiments in science education or how to structure lessons in which new concepts are introduced. We argue that the concept of evidence-based education should be broadened toward a theory-based perspective, in which theories are understood as empirical theories. Teacher education, in turn, should place greater emphasis on developing teachers' theoretical understanding—not by reinforcing a dichotomy between theory and practice, but by demonstrating how theory can fruitfully inform practice and provide powerful tools for reflection.

For the discipline of science education, it is therefore crucial to engage in systematic discussions about the kinds of theories it can offer and how such theories can be developed and empirically examined. As a starting point, we adopt the concept of theory presented in the preceding section to propose a heuristic framework for understanding the role of theory in science education research.

**Consequences for science education research**

In this section, we argue that science education research requires both applied and fundamental research, with each contributing to different forms of theory. We further contend that science education research constitutes a scientific discipline in its own right—one that generates theories originating from, and contributing to, this very field.

*Critical perspectives on role of theory in science education research*

Of course, there are alternative perspectives doubting whether fundamental theories in education can exist, or, in the tradition of Feyerabend (1993), how useful they are. Thomas (2007), for instance, argues that "the allure of theory—and the desire of educators to call their ideas 'theory'—rests historically on its success in other fields, most notably in the natural sciences. […] The domain in which theory has been useful finds no congruence in education" (p. 20). In his view, education comprises too many heterogeneous domains (Gage, 2009, p. 11). Thomas (2007) also contends that theories, in their traditional sense of providing generalizable principles, have almost never emerged from educational research. Gage (2009), however, challenges this perspective, arguing that it neglects "abundant empirical evidence" (p. 21) demonstrating that educational theories—understood as generalizations of human learning and teaching processes—do exist in various forms. Earlier, Gage (1991) questioned whether theories and findings in education merely describe "obvious" results, given that human behavior is rarely unprecedented. Yet he offers numerous counterexamples and concludes: "In any case, the allegation of obviousness may now be countered with the research result that people tend to regard even contradictory research results as obvious. Perhaps even that result will henceforth be regarded as obvious" (p. 16).

Within the field of science education itself, the notion that theories should be central to the research process has been questioned. Geelan (2003), for instance, argues that the dominance of natural science paradigms in educational research may have reached a dead end. The observation that theories have rarely guided educational reform might, in his view, even reflect the impossibility of developing fundamental theories in science education. As he notes, "The long love affair with the methods of the physical sciences as the paradigm of inquiry in education is ending with a sense of frustration" (Geelan, 2003, p. 169). He further contends that science education more closely resembles engineering than the natural sciences and should therefore focus on "the search for powerful, context-sensitive solutions that work in practice rather than for generalizable knowledge about practice" (Geelan, 2003, p. 170). This argument is indeed compelling, particularly his critique of the naïve assumption of a top-down relationship between research and practice—wherein researchers generate theories, teachers are expected to apply them, and any failure to do so is attributed to the practitioners themselves (Geelan, 2003, p. 176).

However, this does not necessarily imply that the development of theories in science education is impossible or undesirable. One could argue that the human tendency to generalize and to seek underlying regularities in observed phenomena is deeply rooted in cognition. Children, for example, begin categorizing experiences and forming abstract concepts early in life—they construct their own "laws of nature" daily, which explains why they enter science classrooms with robust, though often scientifically inaccurate, preconceptions about the physical world (e.g., Duit et al., 2012). We would argue that there is ultimately no alternative to developing scientific theories in educational research. Teachers require frameworks to guide their decisions; and if those frameworks are not grounded in scientific theory, they will inevitably be based on subjective theories or beliefs—often implicit, untested, and naïve. As Geelan (2003, p. 170) notes, "reform movements are frequently grounded in ideology," which underscores the risk of allowing practice to be guided by unexamined assumptions rather than systematic inquiry. That said, theory in science education may indeed arise as a "byproduct" of research (Geelan, 2003, p. 176).

Critiques of theory in education are often rooted in objections to quantitative research, or at least to its dominance (also in Geelan, 2003). As already mentioned above, we fully acknowledge the high value of qualitative research. Even in the natural sciences, a view of research as primarily hypothesis-testing would be one-dimensional (Heering & Höttecke, 2014)—this is even more true for educational research. Science education requires applied research that empowers teachers to address their own pedagogical challenges (Whitehead, 1989). Yet, this form of research is not devoid of theory; rather, it produces local theories grounded in practice. At the same time, fundamental research is equally necessary to develop broader theoretical frameworks that transcend specific contexts and contribute to the cumulative advancement of science education as a discipline. We would argue that it may indeed turn out that theories in science education remain more local than one might ideally hope. We acknowledge that theories comparable in scope and universality to those in physics are unlikely to emerge. It might very well be the case that social systems such as science teaching simply cannot be prepared -- in terms of variable control – in ways comparable to physics systems and this would affect the resulting theories as well. Nevertheless, medicine can again serve as a useful role model. Through case studies, medicine also develops local theories; at the same time, it relies on a range of mid-level theories that address specific domains (e.g., the occurrence of diseases or their symptoms; Thagard, 2000), which nevertheless form an essential part of the discipline's scientific foundation (Hucklenbroich, 2025). Compared to science education, the role of theory in medicine has been discussed lately in much greater detail (Hucklenbroich, 2025). Likewise, science education should continue to strive for increasingly general and coherent theoretical frameworks. There is nothing unscientific about a field aspiring to develop more fundamental theories, even if the most tangible outcomes of this pursuit ultimately lie in the practical problems it helps to solve. While fundamental theories in the strict sense—those encompassing an infinite set of models—may be unattainable, *more fundamental theories* could nevertheless emerge over time. That would mean that as research advances, the number of models subsumed under these theories may gradually increase, thereby expanding their explanatory scope. Theories are beneficial for research. As Gage (2009) sums up hi case against the critique of theory: "It is not true that, in the history of the natural and social sciences, theory has inhibited creativity, originality, flexibility, and intellectual freedom" (Gage, 2009, 39).

***Can there be fundamental theories in science education research?***

But what kind of theory can we expect from science education? Science education is often described as a discipline with many reference fields, including physical sciences, pedagogy, psychology, sociology, and others (e.g., Duit et al., 2012). These reference disciplines provide more fundamental theories—such as cognitive load theory or conceptual change theory—that apply to learning in general, and thus also to science learning. It might, however, be an underestimation of science education to understand it merely as a sub-discipline of psychology or pedagogy. A central question is whether truly fundamental theories can emerge from science education itself, or whether such theories will always remain local.

Gage (2009) argues that learning does not depend on disciplinary content (Gage, 2009, 5). He differentiates the generality of educational theories based on how narrowly they address particular grade levels, subject matters, or student groups. This aligns well with the structuralist view of theory, according to which a theory's degree of fundamentality depends on the scope of its models: the more models a theory encompasses, the more fundamental it becomes. This raises an open question: *Is learning science fundamentally different from learning other subjects?* If so, this is where fundamental theories unique to science education would reside. If not, at least science education researchers are nonetheless indispensable, as they possess expertise in both science and the learning sciences—expertise that cannot be expected from psychologists or pedagogical researchers alone. In this view, the relationship between science education and its reference disciplines could, indeed, be compared to that between engineering and physics: science education research translates theoretical structures from the sciences, informed by learning theories, into instructional practice. The analogy of science education as a form of engineering may ultimately be appropriate and certainly would not limit the discipline. Engineering addresses practical problems and generates new knowledge in the process—knowledge that does not contradict physics but extends beyond what physics alone can provide. One could also argue that human biology is to medicine what physics is to engineering and, maybe, learning psychology to science education.

However, there is a chance that this is too narrow a view of the nature of science education. We acknowledge that few—if any—fundamental theories have been developed in science education so far, which may indicate either that such theories do not exist or that science education research, as a scientific discipline, is not yet sufficiently developed. We would still argue that, in principle, fundamental theories specific to science education could be possible, even if they remain to a certain degree emergent from its reference disciplines as they encompass less models. The key lies in the structure of scientific content: general theories of learning may be fundamental, but they become relevant to science education only when combined with the specific characteristics of scientific subject matter—a feature that may indeed parallel engineering. The natural sciences themselves also provide foundational theoretical frameworks that shape science education. One could regard the content structure of the natural sciences as a moderating variable for theories of pedagogy or learning psychology, in the sense that this structure determines which aspects of these theories can be applied to science teaching and, to some extent, how they can be applied.

We regard it as likely that the synthesis of theories from the sciences and the learning sciences creates a new theoretical quality—one that can be considered *fundamental* within the discipline of science education itself. Science education, thus, generates its own theories rather than merely applying those developed elsewhere. These theories emerge at the intersection of domains, where principles of learning and the epistemic structures of scientific knowledge interact. This perspective aligns with the model of educational reconstruction (Duit et al., 2012) which likewise assumes that integrating the structure of scientific content with learners' conceptual needs is necessary for successful science learning—and that this combination transcends either component in isolation. Combining the content structure of a subject with a theory of teaching does not produce a less generalizable theory, but rather an entirely new *quality* of theory. It is overly simplistic to think of theories according to a "trickle-down" logic, in which a fundamental theory simply subsumes more specific ones.

***There can be fundamental theories in science education research!***

To sum up our argument, theories can be conceived as sets of models: a theory of teaching includes one set of models, a theory of content structure another. When these are combined, a new theory is required—one encompassing both sets of models. That would mean that we can identify theories specific to science education by checking for elements from science combined with elements from learning psychology, pedagogy, or other disciplines concerned with human behavior. Science education may not yet have reached the stage of developing fundamental theories in the strictest sense, but this might just reflect its status as a comparatively young scientific discipline rather than a structural limitation of the field itself.

We would argue that this leads to two main implications for science education:

1. At least one pathway possible for fundamental theories to emerge is for them to begin as local theories inspired by fundamental theories from other disciplines, such as learning psychology, with a very narrow set of models. Subsequent research is then driven by the intention to identify additional applications of the theory and to accumulate further models. This step-wise "theory growth" reflects a typical way in which theories develop within science education.

2. There is no fundamental distinction between qualitative and quantitative research. Although they originate from different research traditions, they do not represent opposing paradigms. Both contribute to the development of theory:

> qualitative research excels in generating (local) theories, whereas quantitative research plays a central role in testing, refining, and extending these theories by determining which models belong to their domain of applicability, thereby helping them evolve into more fundamental theories.

One example may help illustrate these ideas, and the structuralist notion of theories is particularly useful for describing it.

### *An illustrative pathway of theory growth in science education*

There are numerous examples of the pathway of theory growth described above; three are discussed briefly here (the quality criteria of teaching models, the model of educational reconstruction, and activity theory), and one is discussed in more detail (the theory of dialogic explaining).

Gilbert, Boulter, and Rutherford (2000) proposed six criteria for the quality of teaching models, which can be understood as bridges between scientific content and the mental models presented by teachers. Among these criteria are aspects closely related to content structure (e.g., the model should be complete) as well as criteria drawing on psychological considerations (e.g., the model should be coherent). Although this framework is inspired by theories from both the natural sciences and psychology, it becomes applicable only through its integration into a science education context, thereby constituting a science education theory in its own right. Coll and Treagust (2003) built on this framework by investigating students' mental models and extended the theory by introducing an additional model—mental models alongside teaching models. From a structuralist perspective, this development illustrates an example of theory growth: the theory was expanded by incorporating a new model that shares a family resemblance with the existing one.

The model of educational reconstruction (Duit et al., 2012) followed a similar trajectory. It was initially developed as a framework to guide research in science education, based on the idea that students' conceptions of a scientific topic are just as important as the scientific structure of that topic. Over time, it has also been adopted as a framework for planning science instruction; in this sense, the theory has incorporated an additional model. For instructional purposes as well, students' ideas are considered to be just as important as the scientific structure of the content to be taught (Duit, 2015).

Svoboda, Sevian, and Dusen (2025) describe the evolution of activity theory (Engeström, 1999), which also fits the described pathway for theory growth well. It was developed to describe collaborating groups and has grown to include many models e.g., classroom talk or interactive computer-based learning (Svoboda, Sevian, & Dusen, 2025, 3).

To describe the process more in detail, we draw on a theory with which we are familiar from our own research: the theory of dialogic explaining. This theory, developed by Kulgemeyer and Schecker (2009), was originally designed to describe verbal communication between pairs of students in physics classrooms and to explain why some explanations were more likely than others to result in understanding on the part of the explainee. The theoretical foundation is a constructivist understanding of explanation, later elaborated by Kulgemeyer and Geelan (2024): explaining is viewed as a constructivist process of generating explanations that are continuously evaluated for understanding and adapted based on this evaluation. The theory is grounded in two key assumptions: (1) communication in physics must be both subject-adequate (scientifically accurate and complete) and addressee-oriented (adapted to the explainee's needs); and (2) the tools for adaptation include (a) the linguistic level (ranging from scientific jargon to everyday language), (b) the use of examples and analogies, (c) the level of mathematics employed (from qualitative descriptions to formal equations), and (d) the representational means used (e.g., experiments, graphs, photographs). Initially, the theory covered only one model—explaining dialogues between physics students. Over time, additional models were incorporated based on their family resemblance to this original situation: peer tutoring (Kulgemeyer & Schecker, 2013), teacher–student explanations (Kulgemeyer & Riese, 2018; Kulgemeyer et al., 2020), and physics explainer videos (Kulgemeyer, 2018). This has changed the theory by integrating more elements but keeping the structure described above (Kulgemeyer, 2019). This progression demonstrates how a theory in science education can evolve when it is applied and refined across multiple studies and models for the theory are collected. It began as a local theory and gradually accumulated additional models—an evolution that may, in fact, be typical of theories in science education and underscores the value of qualitative research in developing such local theories.

But are these examples—criteria for the quality of teaching models, the model of educational reconstruction, activity theory, and the theory of dialogic explaining—fundamental theories? Are they specific to science education, or merely psychological theories only applied within the field? The model of dialogic explaining integrates elements from science (the requirement of subject adequacy reflects the structure of scientific knowledge) with elements from psychological theory (the adaptation to explainees' needs aligns with Wittwer & Renkl, 2008). The model of educational reconstruction does the same by integrating students' ideas at the same level as science content structure. The criteria for teaching models by Gilbert et al. (2000) is very similar. We would argue that this integration exemplifies how a theory acquires a new quality. For the model of dialogic explaining, the mechanisms of adaptation are inherently science-specific—for instance, the role of mathematics differs markedly from other subjects. It is precisely because of the structure of scientific content that the seemingly straightforward principle of adapting communication becomes useful: simply the claim to adapt is too abstract, means for adaptation are required and provided by the theory. Thus, the theory of dialogic explaining constitutes a science-education-specific theory, emerging from the combination of disciplinary content and learning theory. It represents more than the sum of its components. Nevertheless, it remains an emergent theory rather than a fundamental one. While it may not qualify as a fundamental theory of instruction, it could potentially evolve into a more fundamental theory of science communication through further theoretical integration and empirical expansion. At present, however, it remains an emergent theory, with its models describing explanatory processes primarily in physics education. The same may be true for the model of educational reconstruction, although it has been applied in a far greater number of instructional designs and has therefore accumulated a larger set of models. We would still argue that science education may not yet possess fundamental theories of science learning—but that such theories could emerge through systematic refinement of existing ones.

*Developing theories in science education research*

Both applied and fundamental research are indispensable to find such theories. The starting point for the argumentation is the simple fact that the content of science teaching is subject to democratic control. Curricula depend on what a society considers worth teaching in schools, and there is no fundamental law requiring that science—or any specific content within science—must be included. What is taught changes over time, both due to societal decisions and scientific progress. For this reason, science education research needs fundamental research that is independent of specific content and instead addresses science learning in general. Such research provides fundamental theories in the sense that they can be applied to new, yet unknown topics in science education (based on their family resemblance with existing models for the theory), and are therefore not bound by time-specific or normative curricular factors. At the same time, applied research—most prominently, the development of teaching materials—is equally necessary. Fundamental research, by definition, is not directly applicable, but science education also has a societal mandate to improve classroom practice. Consequently, the development of teaching materials is essential. Ideally, such materials should be theory-driven and empirically tested, thereby both supporting practice and contributing to the testing of theories. Theories that explain why particular teaching materials work are necessarily local in scope, but they are nonetheless essential for directly advancing science teaching, even though they remain subject to change as normative conditions evolve.

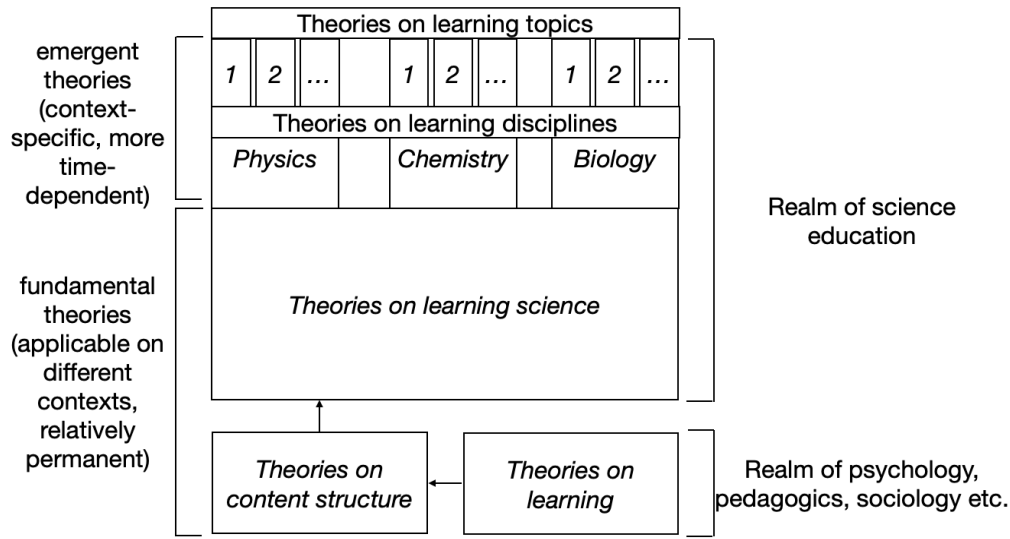

*Figure 3: A heuristic on different kinds of theories in science education*

Figure 3 presents a heuristic illustrating different types of theories in science education. It draws on Gage's (2009, p. 5) notion of hierarchical levels of theory, which aligns well with the structuralist perspective that more fundamental theories encompass a greater number of models than emergent theories. At the same time, it incorporates the idea from Duit et al. (2012) and the model of educational reconstruction that integrating scientific content with the learner's perspective generates an entirely new theoretical quality, rather than a less fundamental one. Given this view, one could argue that theories of content structure should be regarded as moderators between general theories of learning and theories of learning in science, as they determine how learning theories influence the learning of scientific content. In any case, theories of science learning can be fundamental in their own right. A fundamental theory in science education could be part of a larger fundamental theory of education, which would in turn notionally be part of a fundamental theory of human behaviour – but it is still a fundamental theory. As argued above, fundamental research is essential for developing such theories, since they apply across diverse contexts and remain relatively stable even as the content of school science evolves. In contrast, more emergent theories explain in greater detail why learning a specific topic works. They are highly valuable for instructional practice but inherently time-bound, as they lose practical relevance once the specific content they address changes.

What science education research certainly does possess, are theories situated in the upper part of the heuristic shown in Figure 3. These include local theories on, for example, learning Newtonian mechanics or other specific scientific topics, as well as, to some extent, theories on learning physics in general—such as applications of conceptual change theory. These emergent theories are the ones most relevant for teacher education at present, while the search for more fundamental theories of science learning should guide ongoing fundamental research. We would argue that the development of the discipline would benefit if future research consistently sought to elaborate and refine theories, testing them across multiple studies in order to accumulate models. The identification of many such models (that is, applications of a theory) could ultimately elevate a theory to the status of a more fundamental one – like the example of the model of dialogic explaining above.

It may be a bold claim, but we are convinced that science education research has not yet fully embraced this trajectory. Too often, individual studies introduce their own isolated theories, and theoretical frameworks in science education publications tend to be presented merely as "everything we know about a topic" rather than as coherent theories developed cumulatively across studies. Only if researchers systematically build and refine theories—by adding or removing models—will there be a chance for some of these theories to attain the status of fundamental theories. That, however, only works if many studies use the same theory as a background. We therefore argue that future science education research should at least create the opportunity for theories to evolve by striving for greater coherence in the theoretical foundations of research in the discipline, rather than producing a multitude of isolated local theories. We are not bold enough to propose specific candidates for such theories, here. Gage (2009), however, presents remarkable details of such a program for general education. Rather, we suggest that these theories should emerge from existing local theories instead of from seemingly more fundamental theories in general education (see Figure 3). Starting from a single model, additional models can be incorporated based on family resemblance, thereby gradually broadening and refining these theories. The potential for this development certainly exists, given the large number of local theories already present in science education. Advancing science education theories requires the combined contributions of qualitative and quantitative studies, applied and fundamental work, and meta-studies alike.

**Conclusion**

We started the paper with five claims. As a conclusion, we want to summarize the paper by presenting some of the core arguments presented in the paper for these claims. We do not expect all science education researchers to agree with these claims; rather, our aim is to initiate a discussion about the nature of our discipline.

*1. Theories are crucial for science education practice because, contrary to common assumptions, they are often more applicable and informative than meta-studies.*

We argue that while meta-analyses are valuable for synthesizing research, they remain limited to describing what works under which conditions and therefore, alone, cannot provide context-independent guidance for teachers. This limitation may even foster skepticism toward science education research, as overgeneralized findings from meta-studies can conflict with teachers' personal experience and thus widen the perceived theory–practice gap. Theories, by contrast, can predict outcomes across a broader range of contexts—including untested ones—because they define their own domain of applicability through structured sets of models (cases in which the model can be successfully applied to predict outcomes). This makes them a more powerful foundation for informing science instruction. Meta-analyses cannot be discarded and replaced with theories but less reliance on meta-analyses and more reliance on theory might benefit science education practice.

*2. Theories are essential for science education research to mature as a scientific discipline. However, studies in this field only rarely engage in the systematic refinement and extension of existing theories.*

It is, of course, a matter of research paradigms whether one accepts that the progress of a scientific discipline involves the incorporation of an increasing number of models into its theories, thereby allowing these theories to become more fundamental over time. However, we emphasize that very often scientific disciplines evolve through this kind of theoretical development. In science education many studies introduce isolated conceptual frameworks that are neither revisited nor empirically refined. Without a concerted effort to expand and test coherent theories, the field risks remaining fragmented. Although psychology and pedagogy provide many foundational theories, science education occupies a unique intersection between the structure of scientific content and the processes of learning. When the epistemic structures of science are integrated with theories of learning, a new theoretical quality emerges with theories from science serving as moderators for theories on general learning. Theories in science education are, thus, not simply emergent theories of, e.g., psychology but can be fundamental theories in their own right even if they are part of a theory of education. In educational research, it may even be observed that the broader and more fundamental a theory becomes, the less practical guidance it offers.

*3. Science education research too often fails to develop coherent theories that organize phenomena through generalizable principles and allow predictions of educational outcomes. Instead, the term theory is frequently used in a superficial sense—as a collection of study results (in the sense of collective pedagogical content knowledge), methodological guidelines, or conceptual frameworks rather than as a structured explanatory system.*

Drawing on philosophy of science, we contrast this superficial use of "theory" with a structuralist notion that views theories as systematic constructs containing models which define the domains of application. Models can be understood as instances in which a theory has demonstrated its explanatory power; additional models are identified through their family resemblance to existing ones and supported by empirical evidence confirming the predictive power of a theory for the new model. Much of science education research restricts the concept of theory to descriptive frameworks or collective PCK, rather than explanatory systems (cf. Treagust & Harrison, 2000) capable of organizing and predicting phenomena across contexts.

*4. Both fundamental and applied research, with qualitative and quantitative data, are necessary to develop theories in science education.*

Fundamental research provides enduring theories of science learning that remain valid even as curricula evolve. Applied research, for example action research, in contrast, develops local theories that explain why specific instructional approaches work and is essential to fulfill the societal mandate of directly improving science teaching. In principle, local theories can evolve into more fundamental theories as they accumulate additional models. This process is particularly important for science education research because it highlights the complementary value of qualitative research and quantitative research. From a structuralist perspective, qualitative research can be understood as contributing to the description of individual models of a theory, often serving to develop its initial models. Quantitative research, in contrast, contributes to identifying the factors that are relevant across the family of models belonging to a theory, thereby supporting generalization. We suggest that this pathway—from local to fundamental theories—could be especially fruitful for science education; however, it requires systematic series of studies that build upon and refine shared theoretical frameworks.

*5. A more nuanced engagement with theory in science education could substantially enrich current debates on evidence-based education.*

As argued above, even well-conducted meta-analyses are not sufficient to guide decision-making in educational contexts. By positioning theory as the conceptual bridge between meta-analyses and teaching practice, we propose "theory-based education" as a valuable extension of the evidence-based paradigm. Theories can guide both research design and instructional practice, as well as support practitioners in interpreting their experiences through systematic reflection. Of course, this does not stand in opposition to evidence-based education, as theories are always understood as empirical theories that include models based on evidence. A deeper engagement with theory therefore offers a promising pathway to bridging the long-standing theory–practice gap in science education.

ROLE OF THEORY IN SCIENCE EDUCATIONColl, R. & Treagust, D. F. (2003). Investigation of secondary school, undergraduate and graduate learners' mental models of ionic bonding. *Journal of Research in Science Teaching, 40*(5), 464-486. https://doi.org/10.1002/tea.10085

Davies, P. (1999). What is evidence-based education? *British Journal of Educational Studies*, *47*, 108–121.

De Bruyckere, P., Kirschner, P. A., & Hulshof, C. (2020). More urban myths about learning and education: Challenging eduquacks, extraordinary claims, and alternative facts. Routledge.

Deci, E. & Ryan, R. (2008): Self-determination Theory: a macrotheory of human motivation, development, and health. *Canadian Psychology* 49, 182–185.

Duit, R., Gropengießer, H., Kattmann, U., Komorek, M., & Parchmann, I. (2012). The model of educational reconstruction – a framework for Improving teaching and learning science. In: Jorde, D. & Dillon, J. (Eds.) *Science education research and practice in europe*. SensePublishers. https://doi.org/10.1007/978-94-6091-900-8_2

Duit, R. (2015). Model of educational reconstruction. In: Gunstone, R. (Ed.) Encyclopedia of science education. Springer. https://doi.org/10.1007/978-94-007-2150-0_157

Edelsztein, V., & Cormick, C. (2025). Who says scientific laws are not rxplanatory?: on a curious clash between science education and philosophy of science. *Science & Education*, *34*(1), 345–376. https://doi.org/10.1007/s11191-023-00465-0

Edwards, A., Sebba, J., & Rickinson, M. (2007). Working with users: Some implications for educational research. *British Educational Research Journal*, *33*(5), 647–661. https://doi.org/10.1080/01411920701582199

Engeström, Y. (1999). Activity theory and individual and social transformation. In *Perspectives on Activity Theory*, (p.19–38). Cambridge University Press.

Gage, N. (2009). *A conception of teaching*. Springer.
40

ROLE OF THEORY IN SCIENCE EDUCATIONKulgemeyer, C., & Riese, J. (2018). From professional knowledge to professional performance: The impact of CK and PCK on teaching quality in explaining situations. *Journal of Research in Science Teaching*, *55*(10), 1393–1418. https://doi.org/10.1002/tea.21457

Kulgemeyer, C., Borowski, A., Buschhüter, D., Enkrott, P., Kempin, M., Reinhold, P., Riese, J., Schecker, H., Schröder, J., & Vogelsang, C. (2020). Professional knowledge affects action-related skills: The development of preservice physics teachers' explaining skills during a field experience. *Journal of Research in Science Teaching*, *57*(10), 1554–1582. https://doi.org/10.1002/tea.21632

Kulgemeyer, C., & Geelan, D. (2024). Towards a constructivist view of instructional explanations as a core practice of science teachers. *Science Education*, sce.21863. https://doi.org/10.1002/sce.21863

Laudonia, I., Mamlok-Naaman, R., Abels, S. & Eilks, I. (2018). Action research in science education – an analytical review of the literature, *Educational Action Research* 26(3), 480-495, https://doi.org/10.1080/09650792.2017.1358198

Lysenko, L. V., Abrami, P. C., Bernard, R. M., Dagenais, C., & Janosz, M. (2014). Educational research in educational practice: Predictors of Use. *Canadian Journal of Education*, *37*(2), 1–26.

Manhart, K. (1994). Strukturalistische Theorienkonzeption in den Sozialwissenschaften: das Beispiel der Theorie vom transitiven Graphen [The structuralist notion of theory in the social sciences]. *Zeitschrift für Soziologie*, 23(2), 111-128

McAlpine, L., Weston, C., Beauchamp, C., Wiseman, C., & Beauchamp, J. (1999). Monitoring student cues: Tracking student behaviour in order to improve instruction in Higher Education. *The Canadian Journal of Higher Education*, 29(3), 113–144. https://doi.org/10.47678/cjhe.v29i3.183335
43

ROLE OF THEORY IN SCIENCE EDUCATION